\documentclass[fleqn]{ICLASS2018}
\begin{document}

\title{Dynamics of a Spatially Developing Liquid Jet with Slower Coaxial Gas Flow}
\runningheads{14$^{th}$ ICLASS 2018}{Dynamics of Spatially Developing Coaxial Liquid Jet}

\author{A.~Zandian\footnotemark[1], W.~A.~Sirignano}
\address{Department of Mechanical and Aerospace Engineering, University of California, Irvine, CA 92697, USA}
\author{F.~Hussain}
\address{Department of Mechanical Engineering, Texas Tech University, Lubbock, TX 79409, USA}

\footnotetext[1]{\normalsize{Corresponding author:
\href{mailto:azandian@uci.edu}{\textstyleInternetlink{azandian@uci.edu}}}}

\abstract{A three-dimensional round liquid jet within a low-speed coaxial gas flow is numerically simulated and explained via vortex dynamics ($\lambda_2$ analysis). The instabilities on the liquid-gas interface reflect well the vortex interactions around the interface. Certain key features are identified for the first time. Two types of surface deformations are distinguished, which are separated by a large indentation on the jet stem: First, those near the jet start-up cap are encapsulated inside the recirculation zone behind the cap. These deformations are directly related to the dynamics of the growing cap and well explained by the vortices generated there. Second, deformations occurring farther upstream of the cap are mainly driven by the Kelvin-Helmholtz (KH) instability at the interface. Three-dimensional deformations occur in the vortex structures first, and the initially axisymmetric KH vortices deform and lead to several liquid lobes, which stretch first as thinning sheets and then either continue stretching directly into elongated ligaments -- at lower relative velocity -- or perforate to create liquid bridges and holes -- at higher relative velocity. The different scenarios depend on Weber and Reynolds numbers based on the relative gas-liquid velocity as was found in the temporal studies. The deformations in the upstream region are well portrayed in a frame moving with the convective velocity of the liquid jet. The usefullness of the temporal analyses are now established.}

\section{Introduction}

When a liquid jet discharges into a gaseous medium, it becomes unstable and breaks into droplets due to the growth of instabilities. In combustion and jet propulsion applications, the common purpose of breaking a liquid stream into spray is to increase the liquid surface area so that subsequent heat and mass transfer can be increased. Even though the liquid jet breakup has been studied theoretically, experimentally and numerically for more than half a century, the liquid surface deformation mechanisms and its causes are still not satisfactorily understood and categorized at different flow conditions. In this study, a spatially developing round liquid jet with slower coaxial gas flow is analyzed numerically. The main objective here is to examine the interaction of the vortices near the liquid-gas interface, and to see how those interactions vary with gas-to-liquid velocity ratio, and their consequent effects on the surface deformation and growth of instabilities.

Lasheras and Hopfinger \cite{ref:Lasheras} in their review of the liquid jet atomization in a coaxial gas stream, categorized the regimes of liquid jet breakup and showed the effects of gas-to-liquid momentum ratio on those regimes. However, they did not relate those regimes to the dynamics of vortices generated prior to atomization. Shinjo and Umemura \cite{ref:Shinjo} briefly touched upon the axial and radial vortices generated in a round liquid jet atomization process (without coaxial flow) and showed that the orientation of vortices determines the orientation of ligaments created during the primary breakup; however, they mainly focused on the vortices near the jet tip and claimed that the primary breakup is mainly affected by the vortices that are convected upstream from the jet tip, without detailing the vortex interactions. More recently, Jarrahbashi and Sirignano \cite{ref:Dorrin1} and Jarrahbashi et al.~\cite{ref:Dorrin2} studied the details of vortex dynamics in a temporal study of a round liquid jet segment, and showed that the vortices can also form far upstream of the jet cap, independent of the vortices shed behind the cap region. They were able to relate the vortex interactions to the surface deformation, lobe formation and perforation. Later, Zandian et al.~\cite{ref:Arash1, ref:Arash2} extended the vortex dynamics analysis to the atomization of planar liquid sheets. They identified three main atomization regimes with different characteristic length and time scales and unique breakup mechanisms based only on the liquid Reynolds number ($Re_l$) and gas Weber number ($We_g$). They showed that one can understand each breakup mechanism by following the vortex interactions near the gas-liquid interface. Ling et al.~\cite{ref:Ling} also observed the hairpin vortex structures emphasized by Zandian et al.~\cite{ref:Arash2} at the surface of a spatial liquid jet, but failed to explain the details of those vortex interactions.

Here, we perform an analysis similar to Zandian et al.~\cite{ref:Arash2}, but with inclusion of a slow coaxial gas flow and for a spatially developing jet. This study shows the validity of the prior temporal studies and their relevance to a real atomization application.

\section{Numerical Methods}

The three-dimensional Navier-Stokes (NS) with volume-of-fluid (VoF) interface-capturing method yield computational results for the round liquid jet which captures the liquid-gas interface deformations after injection.

The incompressible continuity and Navier-Stokes equations follow
\begin{equation}
\nabla \cdot \textbf{u}=0 ,	\hspace{15pt} \frac{\partial (\rho \textbf{u})}{\partial t} + \nabla \cdot (\rho \textbf{u} \textbf{u})= - \nabla p+\nabla \cdot (2\mu \textbf{D}) - \sigma \kappa \delta (d) \textbf{n},
\label{eqn:incomp continuity}
\end{equation}
where $\textbf{D}$ is the rate of deformation tensor, and $\textbf{u}$ is the velocity field; $p$, $\rho$, and $\mu$ are the pressure, density and dynamic viscosity of the fluid, respectively. The last term in the NS equation is the surface tension force per unit volume, where $\sigma$ is the surface tension coefficient, $\kappa$ is the surface curvature, $\delta (d)$ is the Dirac delta function and $\textbf{n}$ is the unit vector normal to the liquid/gas interface pointing away from the liquid.

Direct numerical simulation is done by using an unsteady three-dimensional finite-volume solver for the NS equations for the round incompressible liquid jet and its coaxial gas stream. A uniform staggered grid is used with the mesh size of $2~\mu m$ and a time step of $5~ns$. A third-order accurate QUICK scheme is used for spatial discretization and the Crank-Nicolson scheme for time marching. The continuity and momentum equations are coupled through the SIMPLE algorithm.

The VoF method developed by Hirt and Nichols \cite{Hirt} captures the liquid-gas interface through time and space. The volume fraction $f$, which represents the volume of liquid phase fraction at each cell, is advected by the velocity field \cite{Hirt}:
\begin{equation}
\frac{\partial f}{\partial t} + \textbf{u} \cdot \nabla f = 0.
\label{eqn:level set}
\end{equation}

The 3D computational domain forms a rectangular box, which is discretized into uniform-sized cells. The domain is initially filled with quiescent gas. The liquid jet of diameter $D=200~\mu m$ is injected from the left boundary at time zero with a constant velocity of $U_l=50~m/s$. The domain size is $15D\times6D\times6D$, in the $x$ (axial), $y$ and $z$ (radial) directions, respectively. The coaxial gas stream fills the rest of the inlet boundary with a constant velocity of $U_g=5$, $10$, and $25~m/s$ resulting in velocity ratios of $\hat{U}=0.1$, $0.2$ and $0.5$, respectively. The other sides are outlet boundaries, where the Lagrangian derivatives of the velocity components are set to zero. Similar axisymmetric (2D) domain is also considered and solved for comparison, where necessary. 

The most important dimensionless groupings in this study are the liquid Reynolds number ($Re_l$), the gas Weber number ($We_g$), and gas-to-liquid density ratio ($\hat{\rho}$), viscosity ratio ($\hat{\mu}$), and velocity ratio ($\hat{U}$), as defined below.
\begin{equation}
Re_l = \frac{\rho_l UR}{\mu_l} , \hspace{10pt}
We_g = \frac{\rho_g U^2 R}{\sigma} , \hspace{10pt}
\hat{\rho} = \frac{\rho_g}{\rho_l} , \hspace{10pt}
\hat{\mu} = \frac{\mu_g}{\mu_l} , \hspace{10pt}
\hat{U} = \frac{U_g}{U_l}.
\end{equation}

The jet radius $R$ is the characteristic length, and for the characteristic velocity the liquid jet velocity $U_l$ is mainly used in the literature. However, our results show that the more relevant characteristic velocity in this problem is the relative velocity of the liquid with respect to gas; i.e.~$U_{r}=U_l-U_g$. The subscripts $l$ and $g$ refer to the liquid and gas, respectively. In this study we mainly focus on $\hat{U}$ effects and for this purpose the three values indicated above are considered. For the other four parameters, the chosen values are kept constant: $Re_l=2000$, $We_g=420$, $\hat{\rho}=0.05$, and $\hat{\mu}=0.01$.

Our goal is to study the vortex dynamics and its influence on the liquid surface dynamics in order to understand breakup mechanisms at different coaxial flow conditions. To this end, the $\lambda_2$ criterion introduced by Jeong and Hussain \cite{ref:Jeong} is used to define a vortex. 

\section{Results and Discussion}

\begin{figure}[t]
\begin{minipage}{40mm}
\begin{center}
\includegraphics[width=3.5cm,keepaspectratio=true]{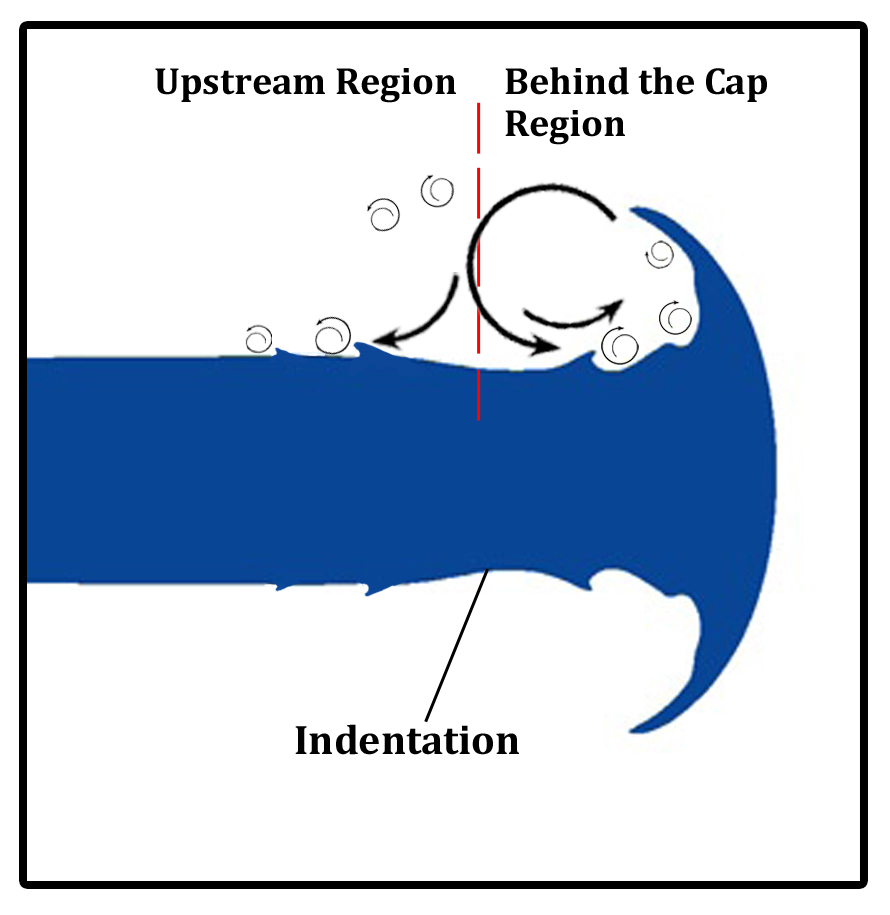}
\caption{Schematic of Vortex regions and wave transmission paths.}
\label{fig:vortex regions}
\end{center}
\end{minipage}
\hfill
\begin{minipage}{45mm}
\begin{center}
\includegraphics[width=4.5cm,keepaspectratio=true]{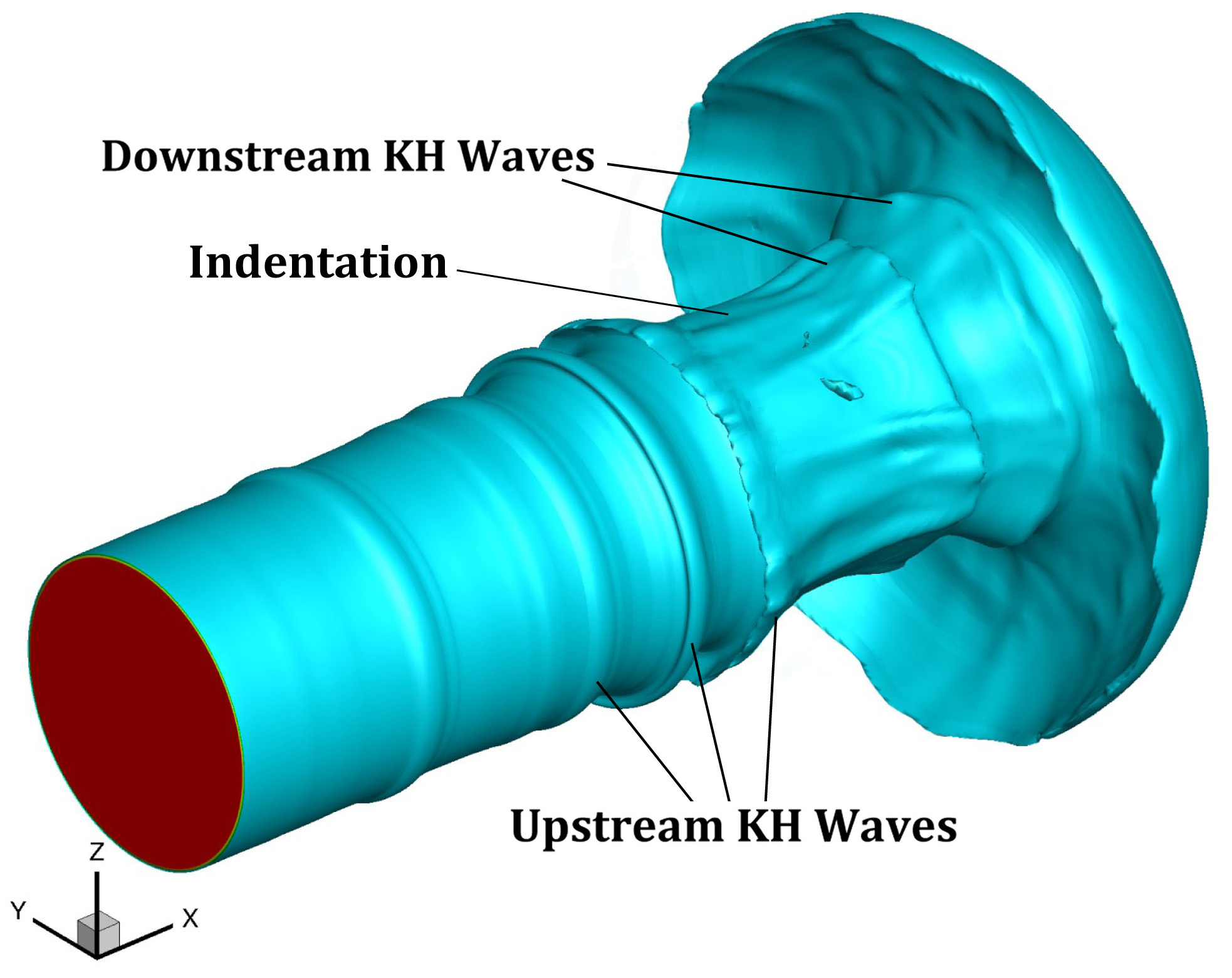}
\caption{Liquid jet surface showing different instability types and their propagation directions; $\hat{U}=0.1$.}
\label{fig:vortex regions 3D}
\end{center}
\end{minipage}
\hfill
\begin{minipage}{65mm}
\begin{center}
\includegraphics[width=6.5cm,keepaspectratio=true]{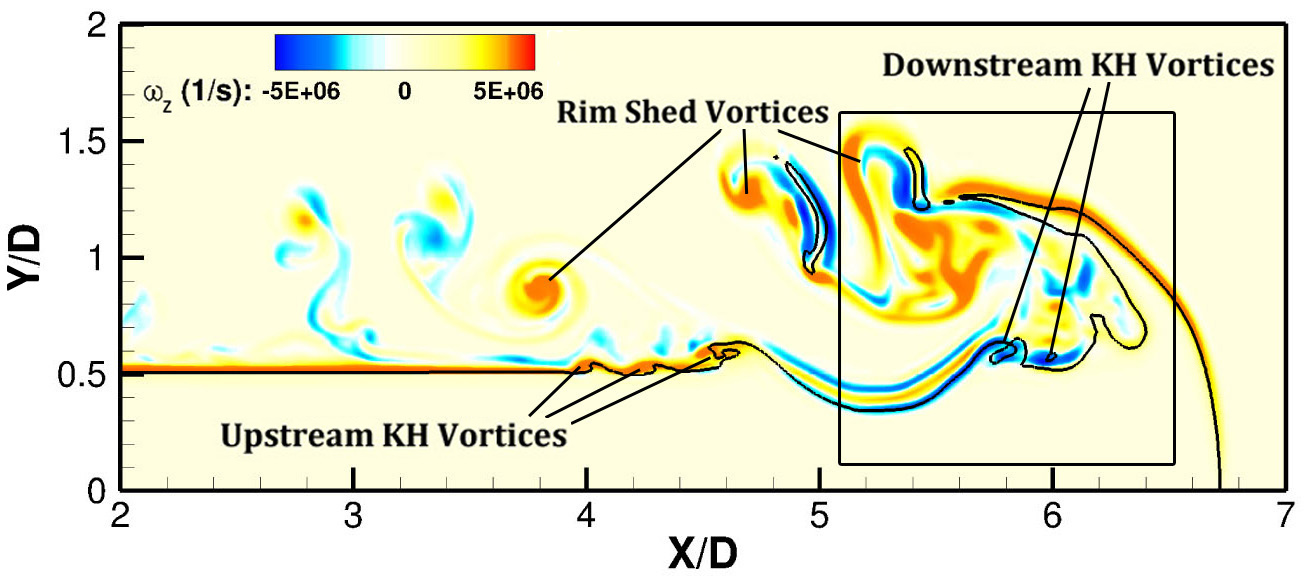}
\caption{Azimuthal vorticity contour ($\omega_z$) and the various vortices generated near the interface in the axisymmetric jet; $\hat{U}=0.1$.}
\label{fig:axisym vorticity}
\end{center}
\end{minipage}
\end{figure}

Two types of surface perturbations are distinguished when a liquid jet is injected in a gaseous medium. These two surface deformations and their regions of occurrence are shown schematically in Fig.~\ref{fig:vortex regions} and the 3D simulation result shown in Fig.~\ref{fig:vortex regions 3D}. The detached droplets and ligaments have been removed in Fig.~\ref{fig:vortex regions 3D} to better display the surface waves on the liquid jet core. The first region is right behind the jet start-up cap, to the right of the broken red line in Fig.~\ref{fig:vortex regions}, and is called the Behind the Cap Region (BCR). In this region, the gas phase velocity is faster compared to the liquid jet and thus, the relative local velocity of the gas stream points downstream. The gas shear creates negative azimuthal vorticity ($\omega_z$) near the interface, which generates KH vortices and consequently downstream facing KH waves, as shown in Fig.~\ref{fig:axisym vorticity}. The vortices in this region are encapsulated inside the recirculation zone behind the cap -- the region inside the box in Fig.~\ref{fig:axisym vorticity} -- and the surface deformations are directly related to the dynamics of the growing cap and can be explained by the vortex interactions in that region. Shinjo and Umemura \cite{ref:Shinjo} briefly discussed this region in their 3D simulation of a round liquid jet with no coaxial flow, and while describing that the vortex dynamics in this region are very complex, they showed that the vortex orientation determines the orientation of the ligaments that are broken from the cap. The surface pattern is not periodic in BCR. Even though this region is not the main focus of our study, kinematics of the cap and the BCR waves are analyzed later in this section.

The second kind of surface deformations occurs farther upstream of the cap, to the left of the broken red line in Fig.~\ref{fig:vortex regions}, and is called the Upstream Region (UR). In this region, the liquid-phase velocity is faster than the gas, and the relative gas velocity points upstream. The gas shear creates positive azimuthal vorticity which creates KH vortices that roll-up the liquid surface and generate upstream facing KH wave pattern. As also mentioned by Shinjo and Umemura \cite{ref:Shinjo}, relatively periodic wave patterns can be observed in this region. They claim that the UR dynamics are highly affected by the BCR dynamics since the shed vortices and the broken droplets are transmitted upstream from the jet cap. However, our analysis shows that the vortices shed from the cap rim mainly affect the droplet propagation in the upstream region and are far from the interface and have minor interactions with the UR KH waves (see Fig.~\ref{fig:axisym vorticity}); thus, the surface dynamics in UR can be studied separately from the BCR in a temporal analysis with periodic boundary conditions similar to studies conducted by Jarrahbashi et al.~\cite{ref:Dorrin1, ref:Dorrin2} and Zandian et al.~\cite{ref:Arash1, ref:Arash2}.  

As shown in Figs.~\ref{fig:vortex regions}--\ref{fig:axisym vorticity}, a large indentation exists between the UR and BCR regions. This indentation is caused by the radially inward gas flow which impinges on the jet stem upstream of the cap. The gas stream then branches into two opposite streams, one flowing downstream and the other upstream in the frame of reference of the jet. The same indentation was also observed in results of Shinjo and Umemura \cite{ref:Shinjo} (Fig.~15a), although not emphasized by them. We draw the line between the two regions at the center of this indentation. Shinjo and Umemura \cite{ref:Shinjo} however, did not identify an exact criterion for the borderline between these two regions, and their BCR and UR regions overlapped at some places and also stretched beyond our proposed segmentation line.

\begin{figure}[b]
	\begin{minipage}{50mm}
		\begin{center}
			\includegraphics[width=5cm,trim={1cm 0.4cm 0.85cm 2.2cm},clip]{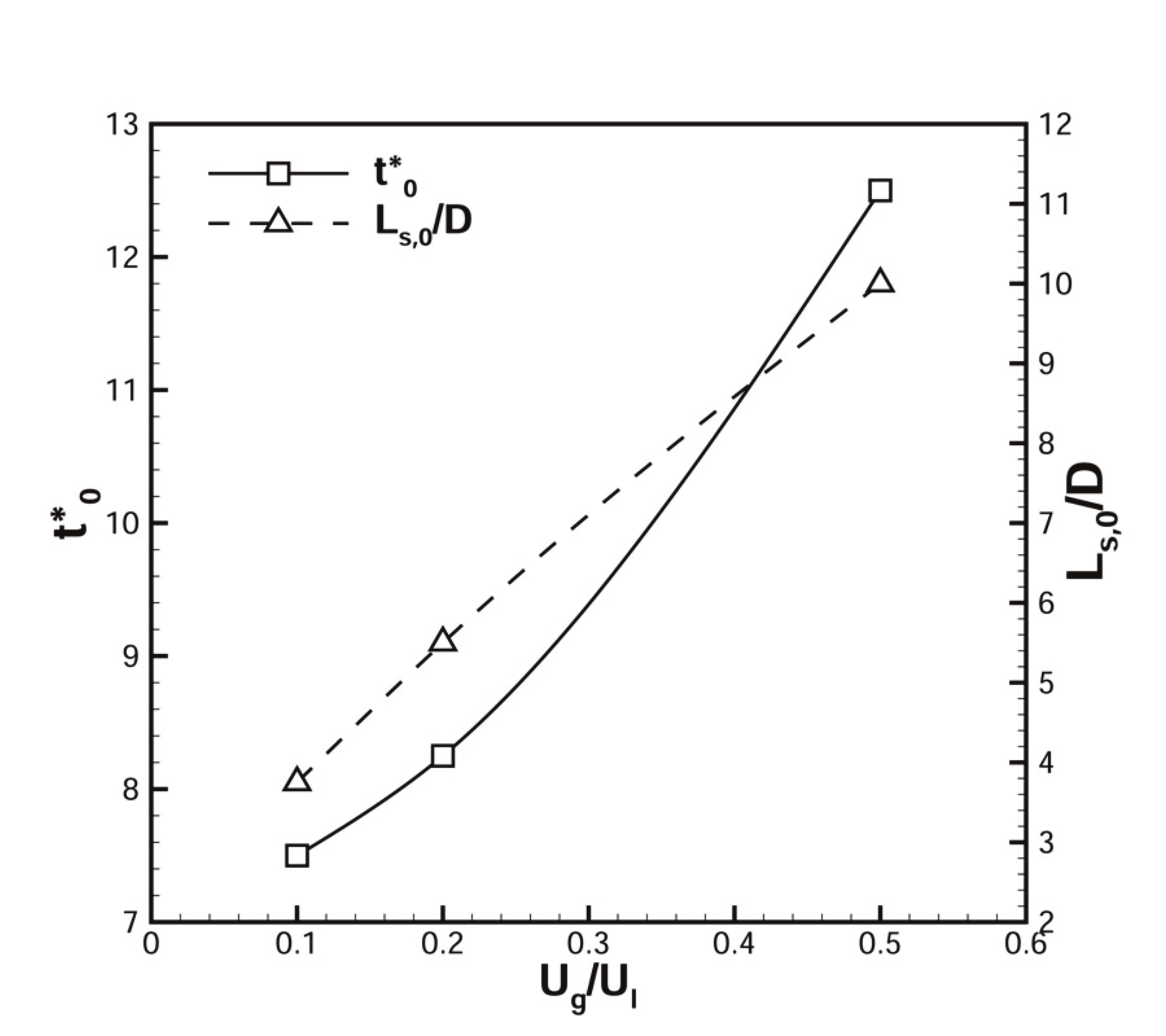}
			\caption{Non-dimensional time and distance of the first perturbation at different velocity ratios.}
			\label{fig:time_length vratio}
		\end{center}
	\end{minipage}
	\hfill
	\begin{minipage}{50mm}
		\begin{center}
			\includegraphics[width=5cm,trim={0.5cm 0.35cm 1.5cm 2.2cm},clip]{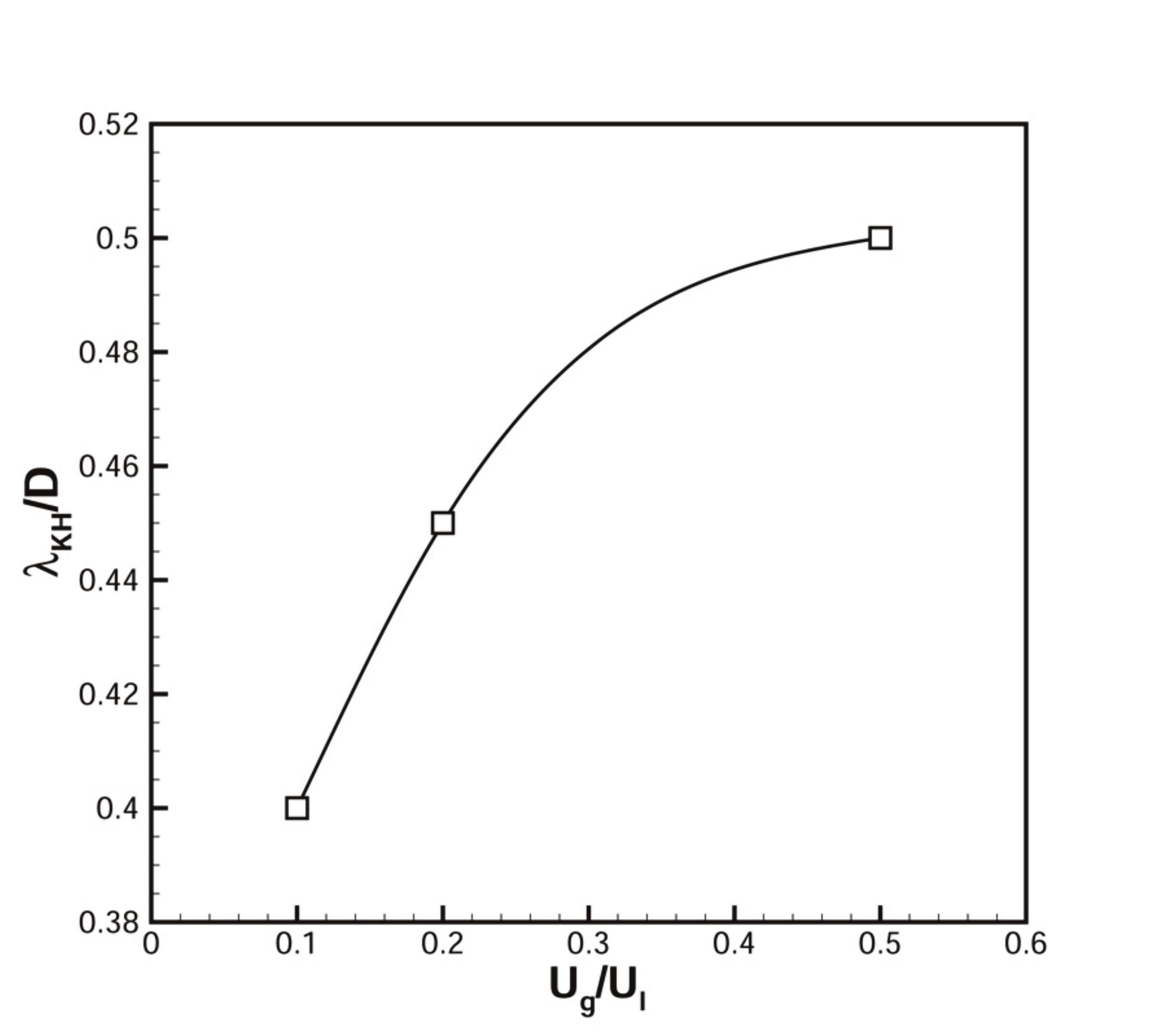}
			\caption{Non-dimensional average KH wavelength for different velocity ratios.}
			\label{fig:lambda vratio}
		\end{center}
	\end{minipage}
	\hfill
	\begin{minipage}{50mm}
		\begin{center}
			\includegraphics[width=5cm,keepaspectratio=true]{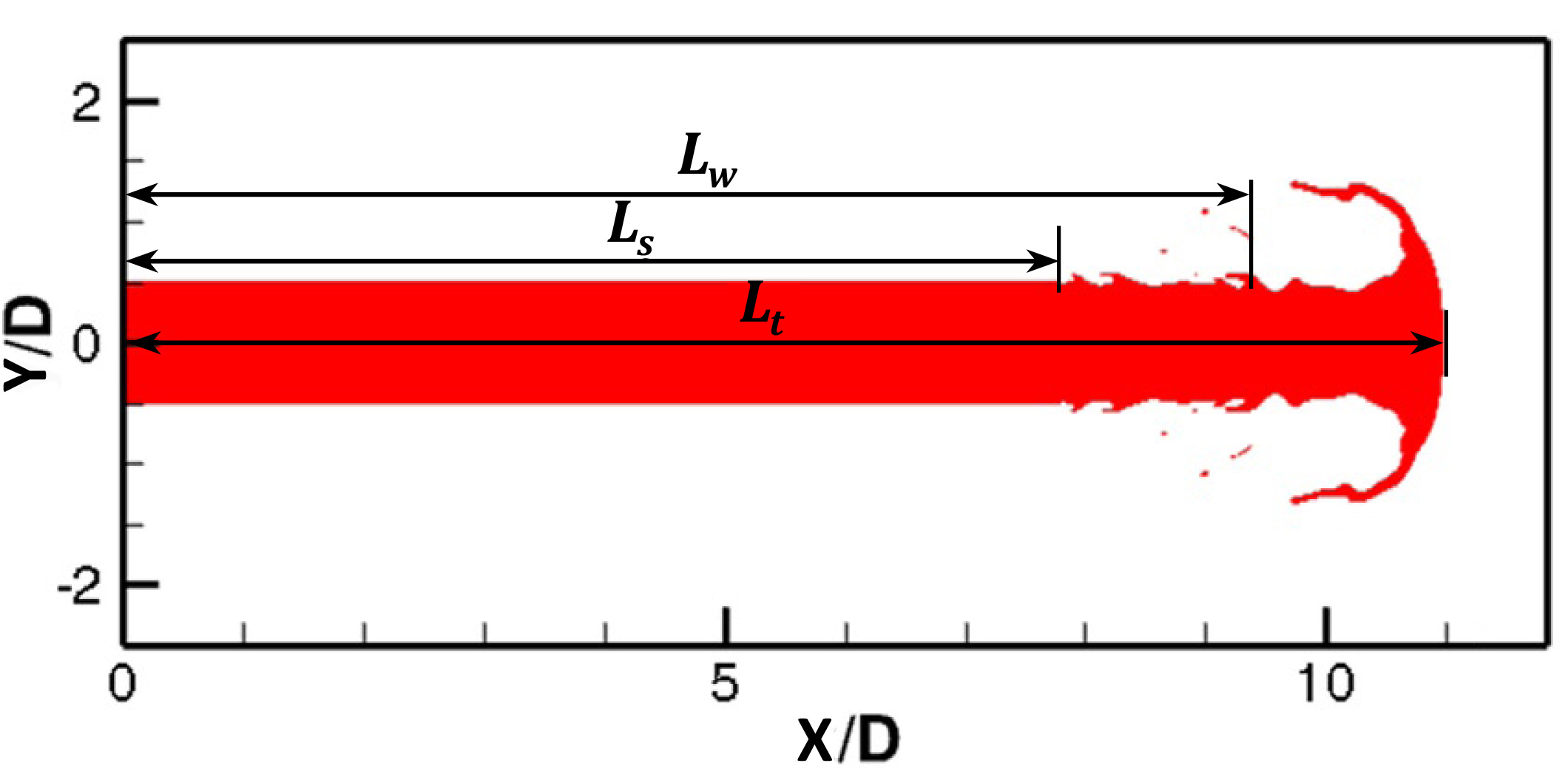}
			\caption{Definition of the tip length ($L_t$), the smooth region length ($L_s$), and the first UR-KH-wave length ($L_w$).}
			\label{fig:length definition}
		\end{center}
	\end{minipage}
\end{figure}

Figure \ref{fig:time_length vratio} shows the time and length of the first KH perturbation occurrence at different velocity ratios. Time has been non-dimensionalized by the injection velocity and jet diameter ($t^*=U_lt/D$), and the length is normalized by the jet diameter. Both jet smooth length and time increase substantially by increasing the coaxial gas velocity. The increase in the length seems to be almost linear, which means that the perturbations are transmitted downstream according to the gas stream velocity. However, the time at which the first perturbation occurs gets delayed exponentially by increasing the velocity ratio.

Figure \ref{fig:lambda vratio} compares the average length of surface waves in the axial direction for different $\hat{U}$. By increasing $\hat{U}$ from 0.1 to 0.5, the average wavelength increases from 80 to 100~$\mu m$. The effect of coaxial gas velocity on axial wavelength becomes less significant as $\hat{U}$ increases. Results of Figs.~\ref{fig:time_length vratio} and \ref{fig:lambda vratio} clearly indicate that the most relevant characteristic velocity in coaxial injection problems is the relative velocity between the liquid and gas streams; i.e.~$U_r$. Thus, since obviously $Re_l$ is the same for all three cases, the most pertinent Reynolds number should also be based on $U_r$ and not the injection velocity as is used in many studies in the literature. As $\hat{U}$ decreases, hence $Re_{l,r}$ increases, axial wavelength decreases, as intuitively expected. 

\begin{figure}[h]
	\begin{minipage}{50mm}
		\begin{center}
			\includegraphics[width=5cm,trim={1.2cm 0.7cm 2cm 2.2cm},clip]{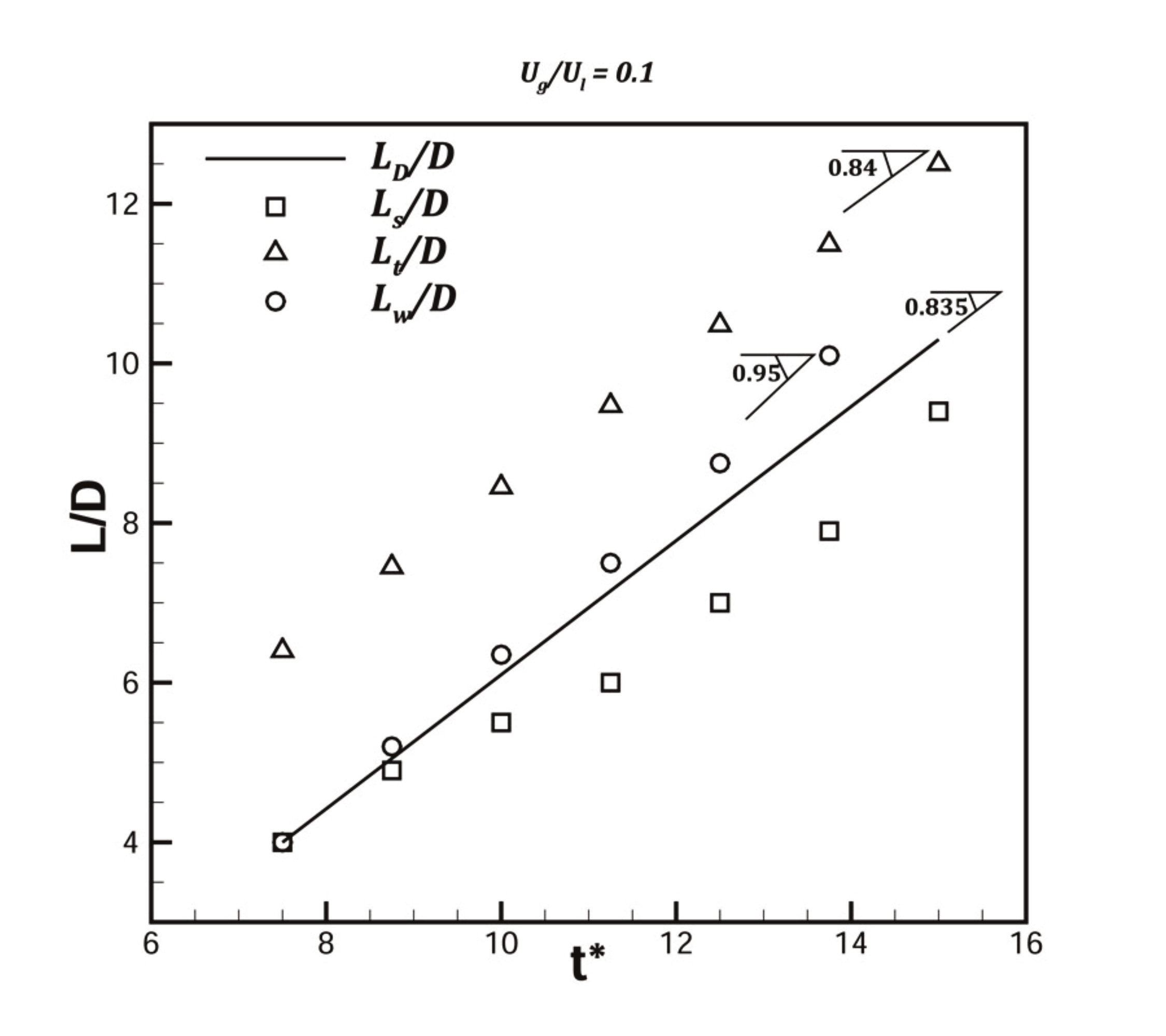}
			\caption{Temporal plot of $L_s/D$, $L_t/D$, and $L_w/D$ for $\hat{U}=0.1$. The solid line indicates simple convection with Dimotakis velocity.}
			\label{fig:vratio01}
		\end{center}
	\end{minipage}
	\hfill
	\begin{minipage}{50mm}
		\begin{center}
			\includegraphics[width=5cm,trim={1.2cm 0.7cm 2cm 2.2cm},clip]{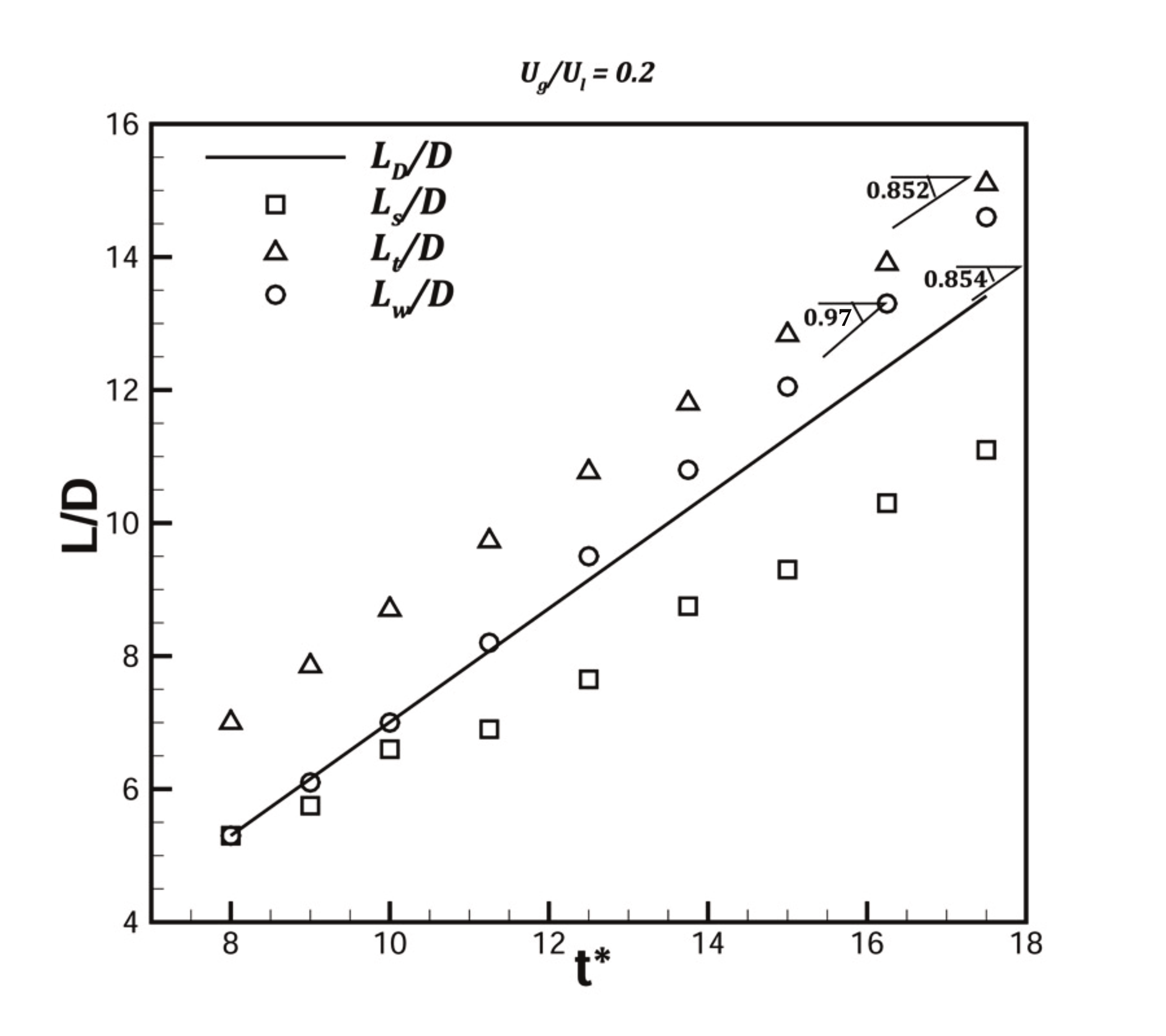}
			\caption{Temporal plot of $L_s/D$, $L_t/D$, and $L_w/D$ for $\hat{U}=0.2$. The solid line indicates simple convection with Dimotakis velocity.}
			\label{fig:vratio02}
		\end{center}
	\end{minipage}
	\hfill
	\begin{minipage}{50mm}
		\begin{center}
			\includegraphics[width=5cm,trim={1.2cm 0.7cm 2cm 2.2cm},clip]{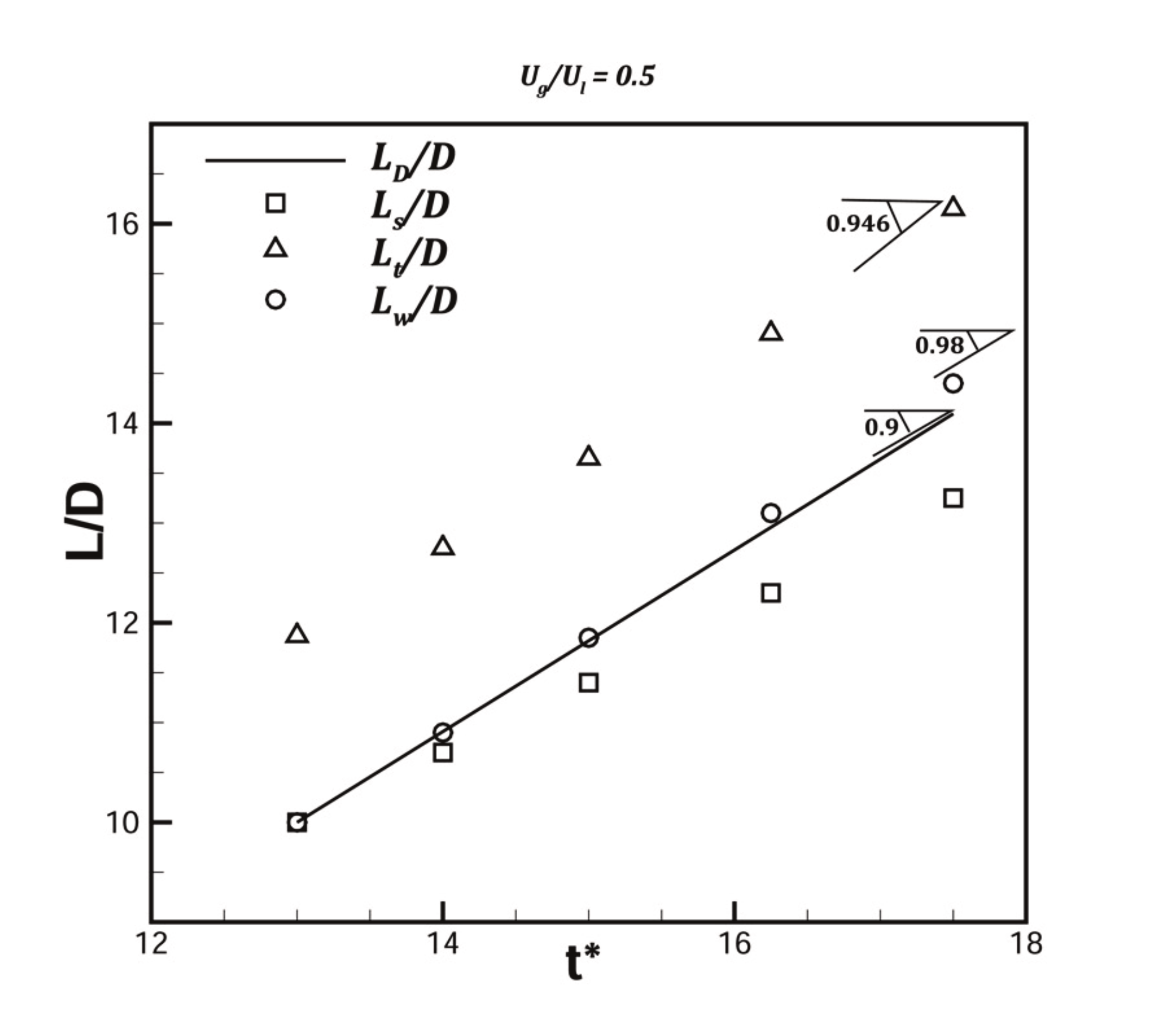}
			\caption{Temporal plot of $L_s/D$, $L_t/D$, and $L_w/D$ for $\hat{U}=0.5$. The solid line indicates simple convection with Dimotakis velocity.}
			\label{fig:vratio05}
		\end{center}
	\end{minipage}
\end{figure}

Figure \ref{fig:length definition} schematically defines the length of the jet tip ($L_t$), the jet smooth length ($L_s$), and the length of the first surface wave in the UR region ($L_w$), which are measured for three $\hat{U}$ cases in time in Figs.~\ref{fig:vratio01}--\ref{fig:vratio05}. All of the lengths are measured from the injection plane and all are normalized by the jet diameter. The average velocity of each length (in terms of the fraction of $U_l$), measured from the slope of each trend, is also computed and presented on the plots. The solid lines indicate the simple convection of the first perturbation with the Dimotakis velocity $U_d=(U_l+\sqrt{\hat{\rho}}U_g)/(1+\sqrt{\hat{\rho}})$.

In all cases, the wave speed follows the Dimotakis speed at early times after its appearance, but it diverges and becomes slightly larger at later times. This divergence is more apparent at lower $\hat{U}$. The smooth length also grows in time for all cases, but it is always below $L_w$. Shinjo and Umemura \cite{ref:Shinjo} also observed a similar difference in the rate of smooth region growth and the injection speed. They concluded that this difference means that the size of the region of influence of the tip is spreading toward upstream as time passes. However, we do not see a direct connection between these perturbations and the vortices generated in the BCR, and the only conclusion that can be drawn here is that new instabilities keep forming upstream of the initial perturbation, resulting in growth of the UR. We showed earlier that UR is not much affected by the BCR and thus, the only reason for the upstream spreading of the instabilities is concluded to be the increase in the strain rate on the surface after the growth of the former KH waves. This triggers new KH vortices, which result in new waves upstream. The rate of growth of the smooth length increases as $\hat{U}$ increases.

\begin{figure}[b]
	\centering\includegraphics[width=\linewidth]{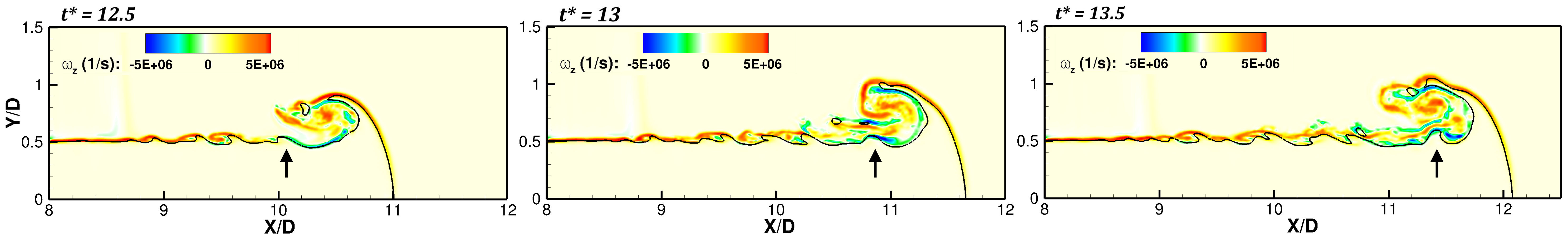}
	\caption{Change in roll-up direction of the KH wave as it moves from UR to BCR region; $\hat{U}=0.2$.}
	\label{fig:direction}
\end{figure}

The tip velocity is always smaller than the wave speed. This means that the KH waves that form in UR, finally enter BCR and catch up with the tip. The rate at which the KH waves reach the tip is directly proportional to the difference between the tip speed and the wave speed. This velocity difference becomes smaller at higher $\hat{U}$ and thus, it takes more time for the UR waves to merge with the cap and disappear. This happens at $t^*=15$ for $\hat{U}=0.1$, and at $t^*=17.5$ for $\hat{U}=0.2$. For $\hat{U}=0.5$, this catchup is so slow that the cap moves out of the domain before the KH wave has enough time to reach it. When the UR waves get into BCR, their roll-up direction changes since the relative gas stream direction changes. This phenomenon is clearly shown in Fig.~\ref{fig:direction}, where the KH wave indicated by the black arrow faces upstream at $t^*=12.5$ while it is in UR, becomes neutral as it enters BCR boundary at $t^*=13$, and faces downstream while it is in BCR at $t^*=13.5$. Notice the formation of a negative KH vortex just as the wave enters the BCR region. 

\begin{figure}[h]
	\begin{minipage}{100mm}
		\begin{center}
			\includegraphics[width=9.9cm]{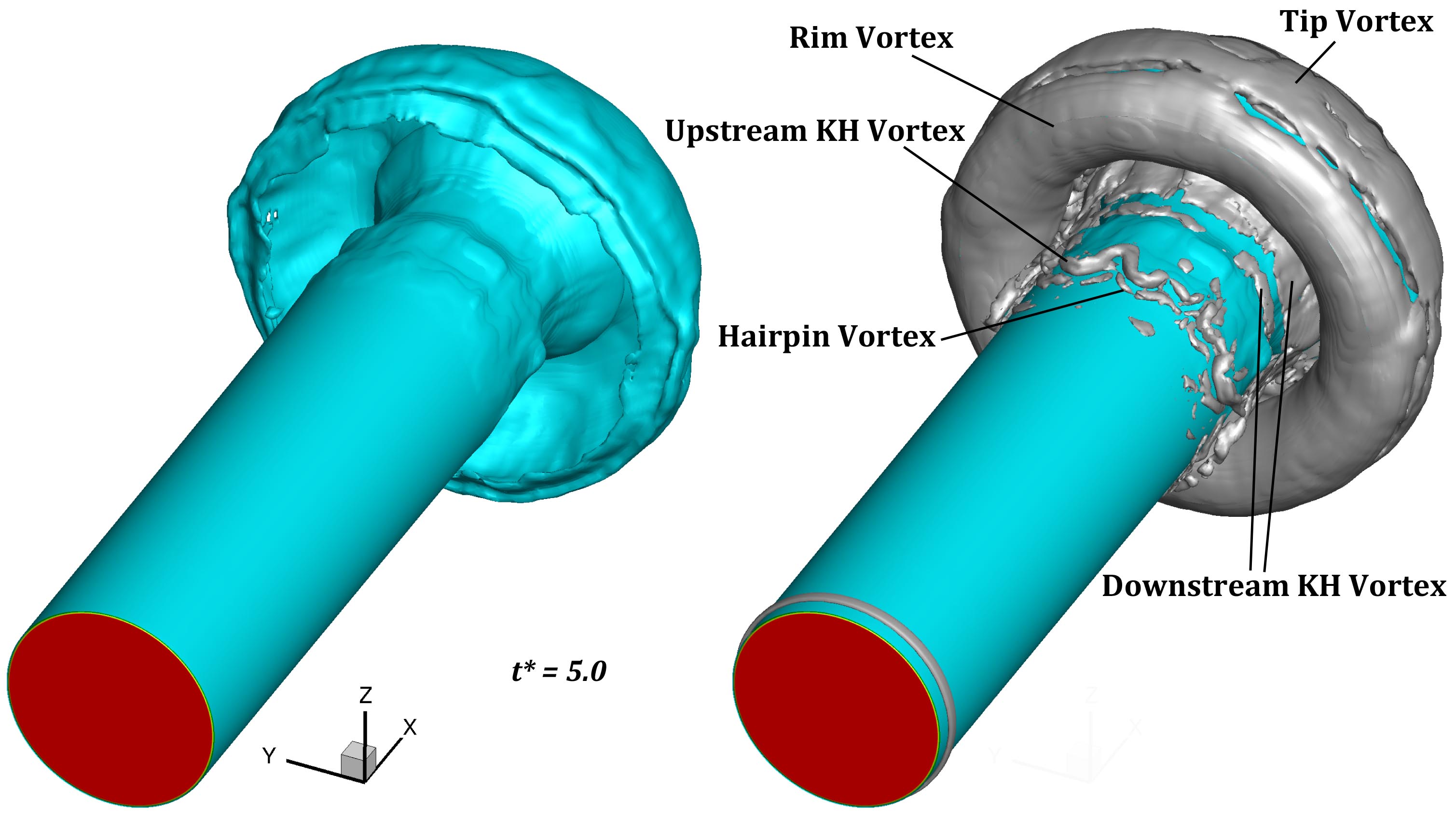}
			\caption{Liquid-jet surface on the left and vortices indicated by $\lambda_2=-10^{11}~s^{-2}$ isosurface on the right at $t^*=5$; $\hat{U}=0.2$.}
			\label{fig:vortex types}
		\end{center}
	\end{minipage}
	\hfill
	\begin{minipage}{50mm}
		\begin{center}
			\includegraphics[width=5cm]{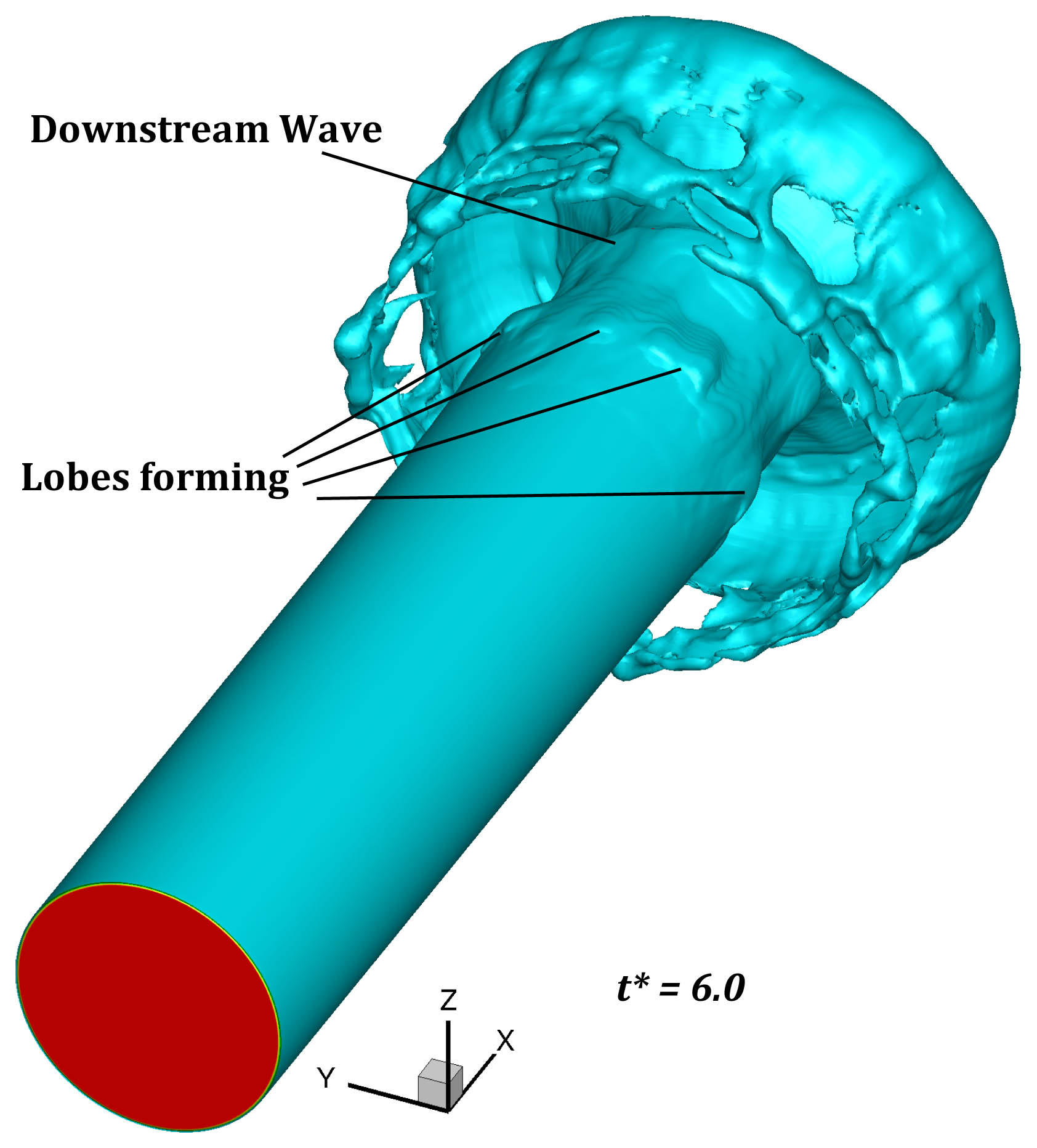}
			\caption{Liquid jet surface at $t^*=6$; $\hat{U}=0.2$.}
			\label{fig:Lobe formation}
		\end{center}
	\end{minipage}
\end{figure}

The liquid jet at a few time steps before the start of perturbations is shown in Fig.~\ref{fig:vortex types}. The vortex structures indicated by the $\lambda_2$ isosurface are also shown in the same figure at the same time step. The vortices that are attached to the jet tip include the vortex structure that covers the front of the jet cap (Tip Vortex), which is caused by the gas flow that goes around the mushroom-shaped cap, and a vortex ring that is shed from the rim of the cap (Rim Vortex), due to flapping of the rim. As discussed earlier, on the stem of the jet, there are two sets of oppositely oriented KH vortex rings which are formed due to the shear caused by the entrained gas in the downstream and upstream directions. As seen in Fig.~\ref{fig:vortex types}, the KH vortices deform and take a hairpin structure before the surface of the jet is deformed. Figure \ref{fig:Lobe formation}, shows that the first liquid lobes are formed at the exact same location where the KH vortices are turned streamwise and create a hairpin structure, at a later time. This conveys that the vortex dynamics drives the surface dynamics, as was first identified by Jarrahbashi et al.~\cite{ref:Dorrin2} and Zandian et al.~\cite{ref:Arash2}. Even though both those studies were temporal with periodic conditions on a liquid segment, our spatial results show that the temporal study of vortex dynamics can capture well the mechanisms in the UR region, where fairly similar physical behaviors occur.

\begin{figure}[b]
	\centering\includegraphics[width=0.8\linewidth]{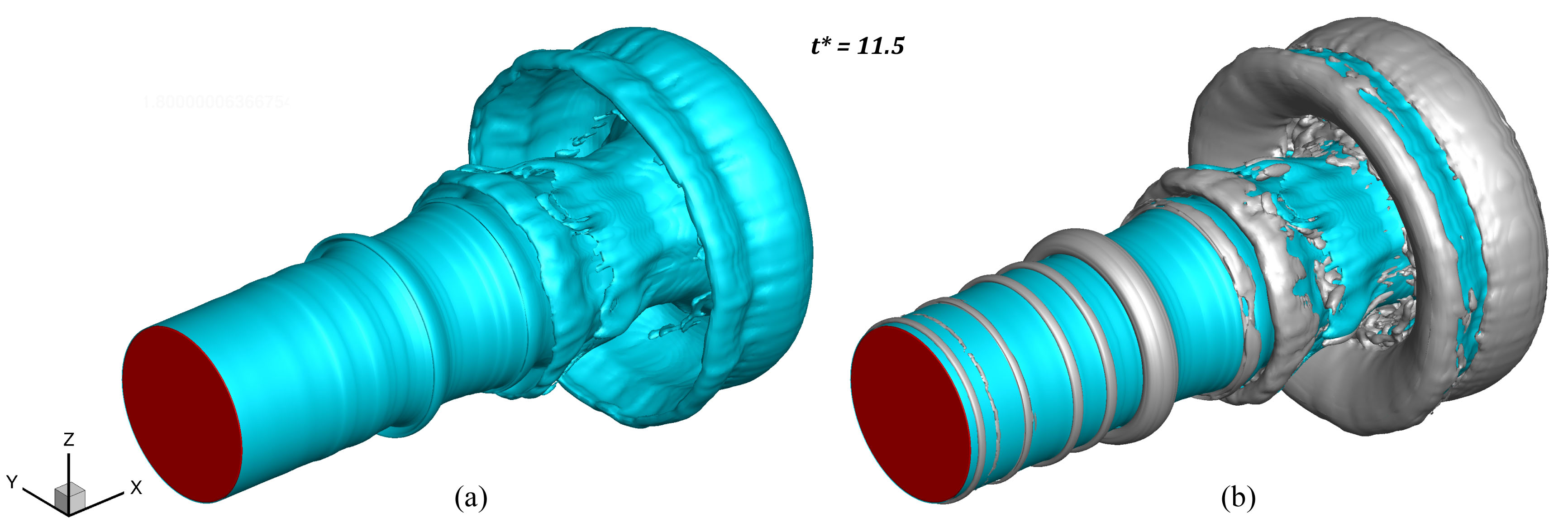}
	\caption{Liquid-jet surface (a), and vortex structures indicated by $\lambda_2=-10^{11}~s^{-2}$ isosurface (b) at $t^*=11.5$; $\hat{U}=0.1$.}
	\label{fig:vortex t11.5}
\end{figure}

Figure \ref{fig:vortex t11.5} shows the liquid jet and its vortices at $t^*=11.5$ for $\hat{U}=0.1$. The KH vortex rings start from an axisymmetric form and grow and deform as they move downstream. Following this change in the KH vortex structure, we can see that the initially axisymmetric KH waves (closer to the nozzle) also become more corrugated as they move downstream. The mode number (azimuthal wavelength) of the lobes is directly related to the number of counter-rotating streamwise vortex pairs that form as the KH vortices become streamwise oriented. Thus, a detailed analysis of the causes of streamwise vortex generation and its growth rate (similar to analyses of Jarrahbashi and Sirignano~\cite{ref:Dorrin1} for round jets and Zandian et al.~\cite{ref:Arash2} for planar sheets) can explain a lot regarding the future behavior of the lobes and their breakup mechanism and size of ligaments and droplets. This analysis will not be quantified here, but the effects of velocity ratio on the mode number will be discussed qualitatively here. 

\begin{figure}[h]
	\begin{minipage}{75mm}
		\begin{center}
			\includegraphics[width=7.5cm]{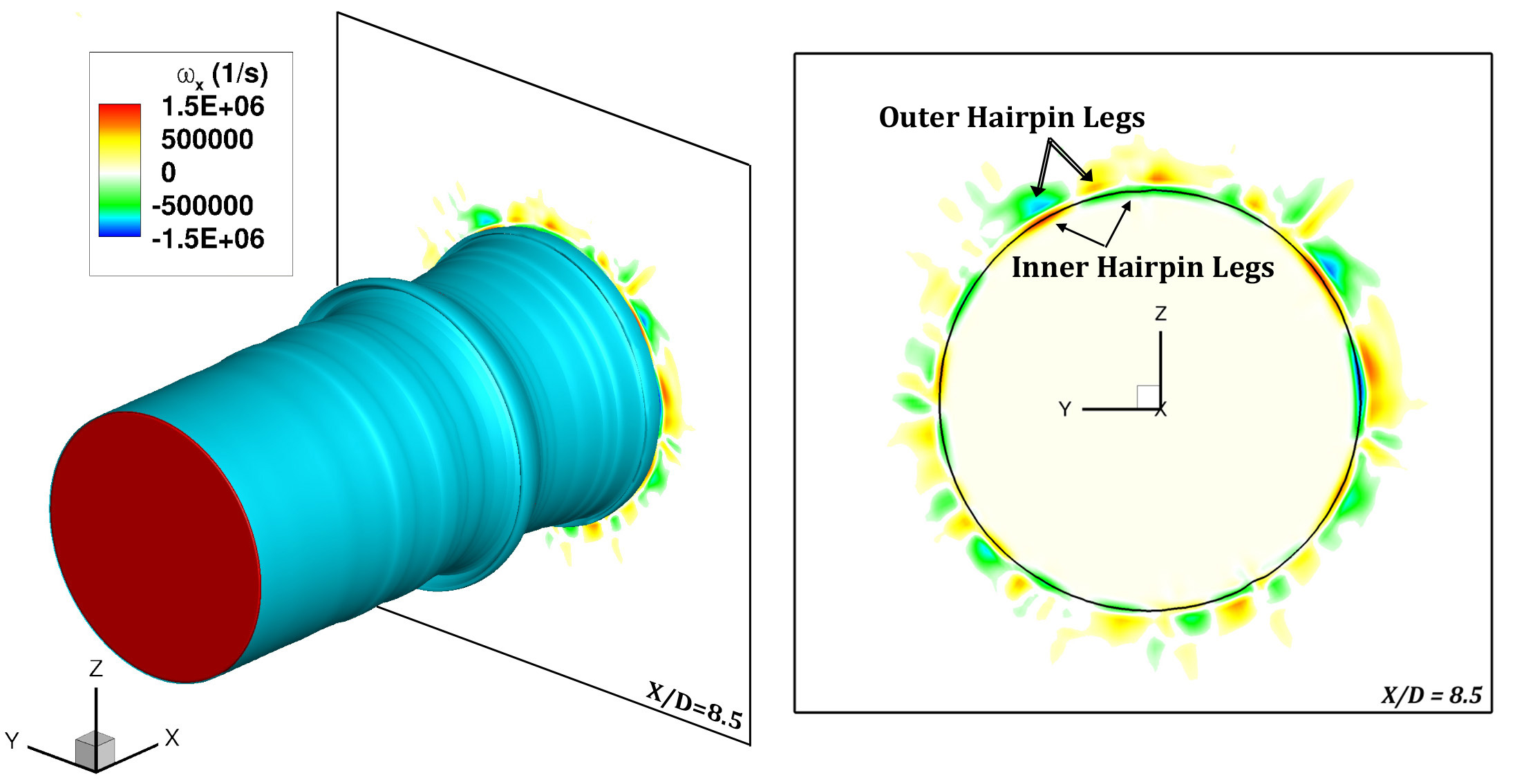}
			\caption{Liquid-jet surface and the axial vorticity ($\omega_x$) contours on the plane intersecting the jet at $x/D=8.5$ at $t^*=11.5$; $\hat{U}=0.1$.}
			\label{fig:OmegaX x/d=8.5}
		\end{center}
	\end{minipage}
	\hfill
	\begin{minipage}{75mm}
		\begin{center}
			\includegraphics[width=7.5cm]{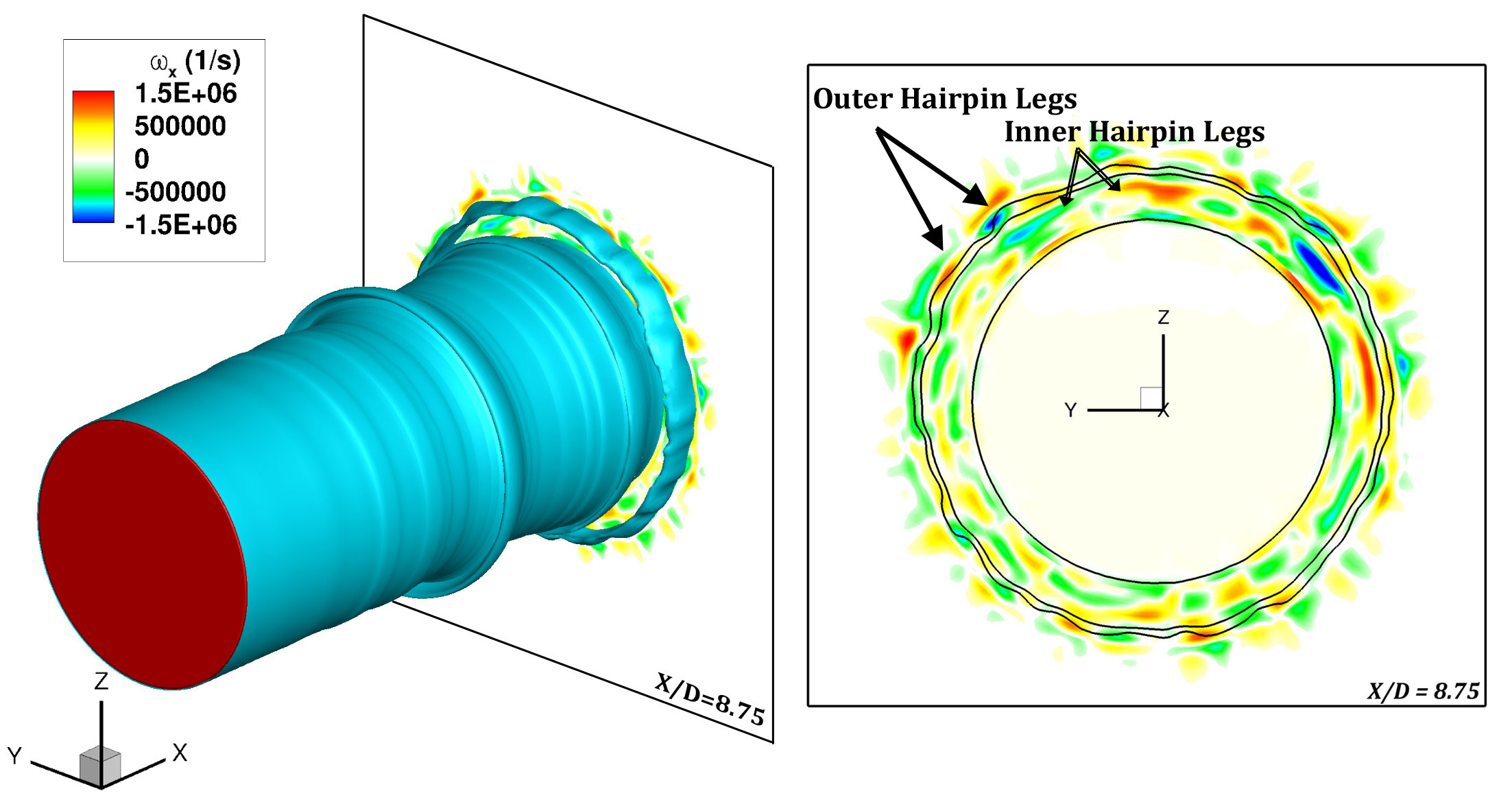}
			\caption{Liquid-jet surface and the axial vorticity ($\omega_x$) contours on the plane intersecting the jet at $x/D=8.75$ at $t^*=11.5$; $\hat{U}=0.1$.}
			\label{fig:OmegaX x/d=8.75}
		\end{center}
	\end{minipage}
\end{figure}

The axial vorticity ($\omega_x$) contours on a spanwise plane intersecting with the liquid jet stem at $x/D=8.5$ and $x/D=8.75$ at $t^*=11.5$ (same time as Fig.~\ref{fig:vortex t11.5}) are shown in Figs.~\ref{fig:OmegaX x/d=8.5} and \ref{fig:OmegaX x/d=8.75}, respectively. The plane in Fig.~\ref{fig:OmegaX x/d=8.5} cuts the jet at the braid of one of the newer KH waves slightly upstream of the former wave that is located at $x/D=8.75$ and is the subject of Fig.~\ref{fig:OmegaX x/d=8.75}. Two layers of counter-rotating streamwise vorticity are seen in Fig.~\ref{fig:OmegaX x/d=8.5}. The inner layer closer to the liquid surface is the hairpin vortex ring with several counter-rotating axial vortices, a pair of which is indicated by the simple arrows. This hairpin stretches upstream and over the next consecutive KH wave upstream of the current wave. Right on the outer side of this vorticity layer is another layer of counter-rotating vorticity, shown by double-lined arrows. Since these counter-rotating vortex pairs are $180^\circ$ out of phase with respect to the inner layer, it is concluded that this layer belongs to another hairpin vortex layer with opposite direction. This layer is called the outer hairpin layer and is stretched downstream and underneath the next downstream KH wave, shown in Fig.~\ref{fig:OmegaX x/d=8.75}. The reason why much axial deflection is still not seen in this vortex ring in Fig.~\ref{fig:vortex t11.5} is that the $\omega_x$ magnitude is an order of magnitude smaller than the azimuthal vorticity ($\omega_\theta$) magnitude at this time. Even though the vorticity layers are not very neat and organized at all azimuthal locations of this picture, seven counter-rotating vortex pairs are distinguished in Fig.~\ref{fig:OmegaX x/d=8.5}, which indicates that seven lobes are expected to form on this KH wave later.

In the downstream cross-sectional plane shown in Fig.~\ref{fig:OmegaX x/d=8.75}, three counter-rotating axial vortices are observed. The outer layer -- indicated by thick arrows -- is the outer hairpin vortex layer for this new wave which stretches downstream. This hairpin layer is right on the outer surface of the KH wave. Right on the inner side of this wave, there is another layer of counter-rotating hairpin vortex, indicated as inner hairpin layer. This layer is the same outer hairpin layer seen in Fig.~\ref{fig:OmegaX x/d=8.5} and is stretched underneath the wave shown in Fig.~\ref{fig:OmegaX x/d=8.75}. A comparison between the counter-rotating vortex pairs of these two layers shows that they both belong to the same hairpin structure that wraps over the upstream wave and under the next downstream wave. There is another vortex ring on the inner side of this hairpin, which is less organized and more chaotic, but with smaller axial vorticity component. This layer is part of the KH vortex located underneath the wave and slightly deflected. 

\begin{figure}[t]
	\begin{minipage}{75mm}
		\begin{center}
			\includegraphics[width=7.5cm]{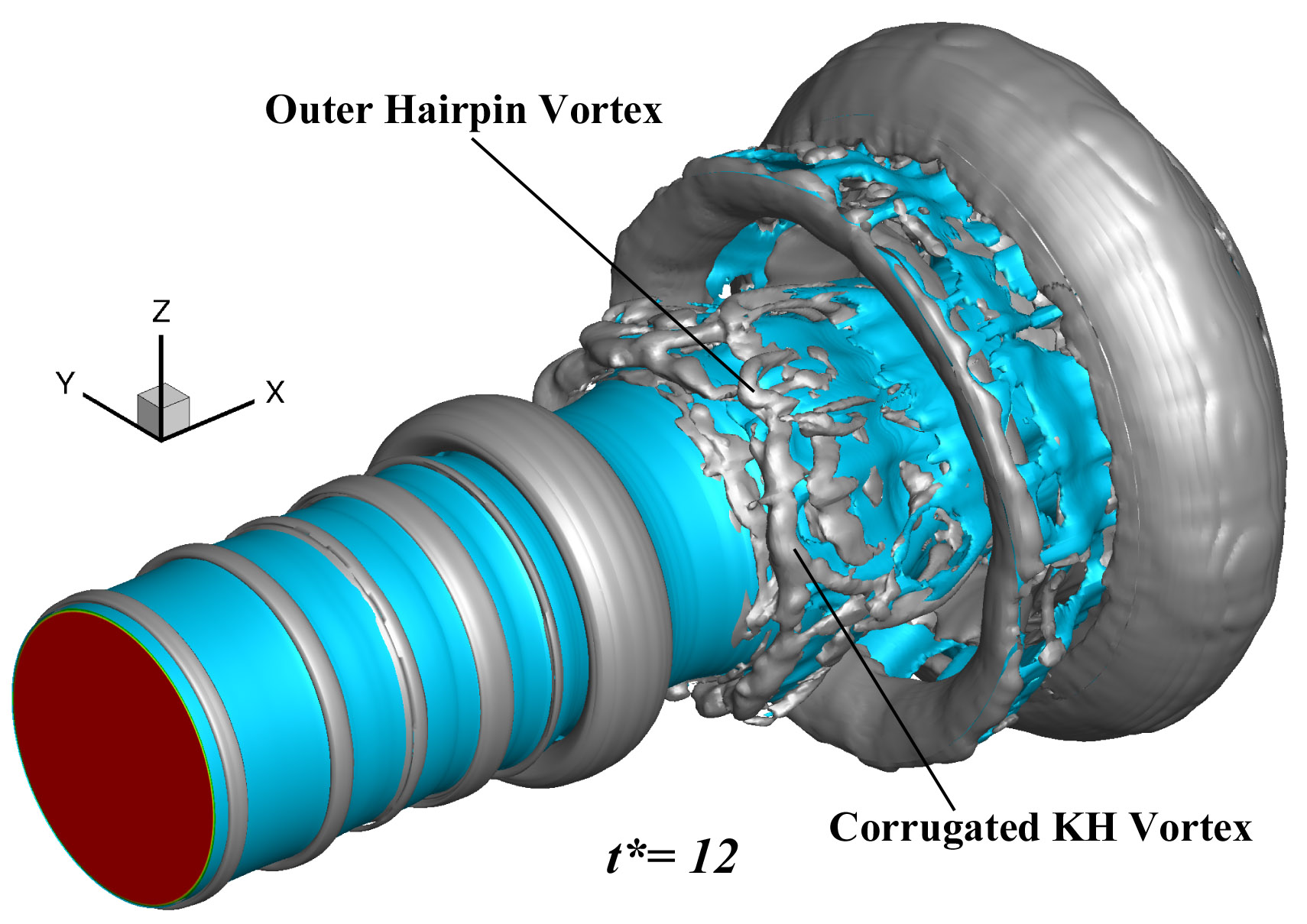}
			\caption{Liquid-jet surface (blue) and vortex structures (gray) at $t^*=12$; $\hat{U}=0.1$.}
			\label{fig:vortex t12}
		\end{center}
	\end{minipage}
	\hfill
	\begin{minipage}{75mm}
		\begin{center}
			\includegraphics[width=7.5cm]{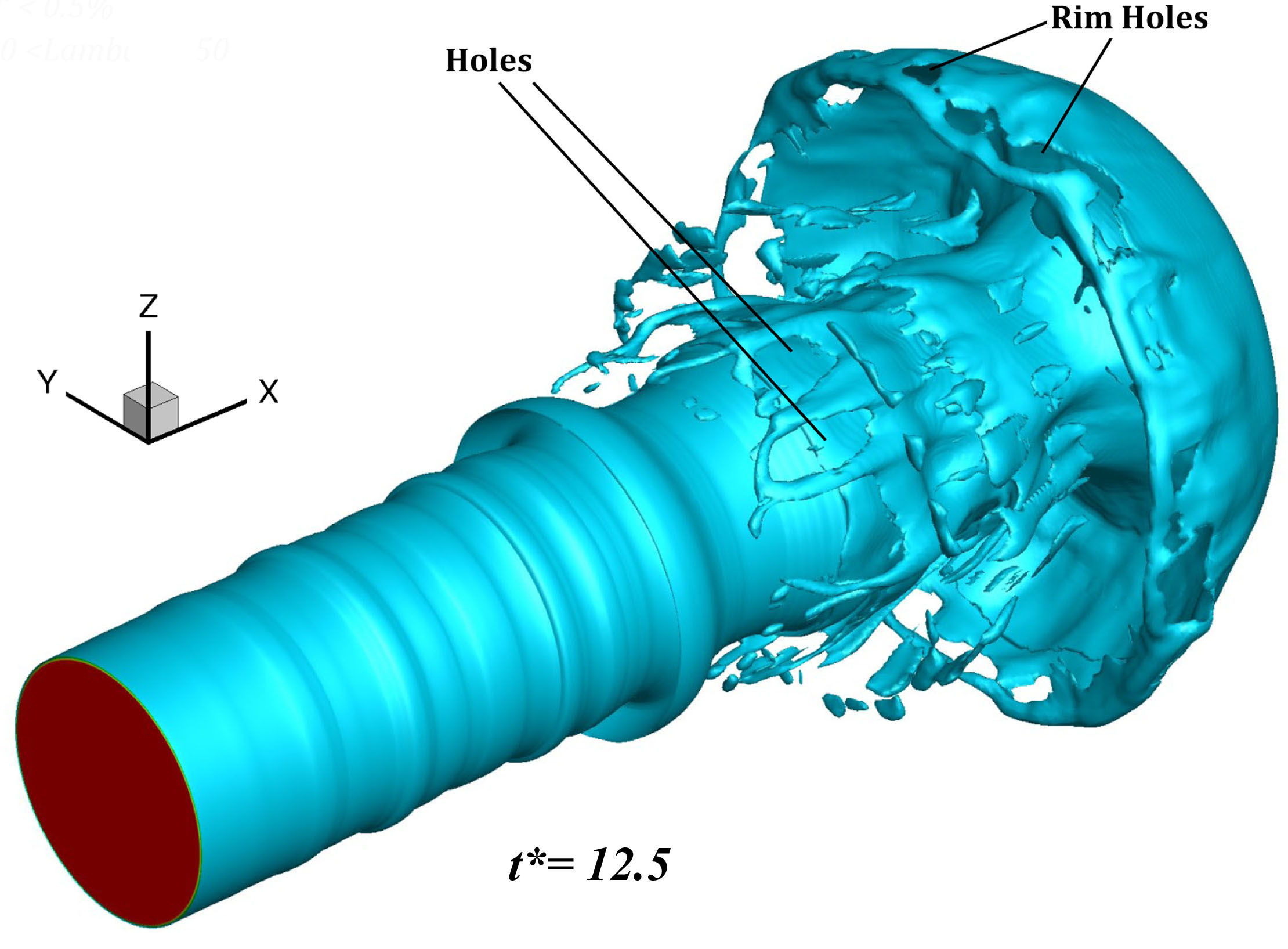}
			\caption{Liquid-jet surface at $t^*=12.5$; $\hat{U}=0.1$.}
			\label{fig:jet t12.5}
		\end{center}
	\end{minipage}
%	\hfill
%	\begin{minipage}{50mm}
%		\begin{center}
%			\includegraphics[width=5cm]{Uratio01_surface_t34.jpg}
%			\caption{Liquid-jet surface at $t^*=15.5$; $\hat{U}=0.1$.}
%			\label{fig:jet t15.5}
%		\end{center}
%	\end{minipage}
\end{figure}

The effects of the counter-rotating axial vortex pairs shown in Figs.~\ref{fig:OmegaX x/d=8.5} and \ref{fig:OmegaX x/d=8.75} are more clear at a later time ($t^*=12$) shown in Fig.~\ref{fig:vortex t12}. When $\omega_x$ grows enough to become comparable to $\omega_\theta$, three-dimensional instabilities occur and the vortices lose their axisymmetry. This phenomenon creates corrugations in the KH vortex ring and also larger axial stretch on the hairpin vortices that are also stretched by the KH vortex. The corrugated KH vortex and the hairpin vortex that stretches over it are shown in Fig.~\ref{fig:vortex t12}. The inner hairpin vortex is not clearly seen in this figure since the KH vortex and the liquid lobes on the outer side of this hairpin block those hairpins from the view. As shown by Zandian et al.~\cite{ref:Arash2}, overlapping of these oppositely oriented counter-rotating hairpins that are on the outer and inner sides of the lobe, thins the lobe at its center and creates holes on the lobes. Thinning of the lobes (wave) can be clearly seen in the cross-sectional view of the plane illustrated in Fig.~\ref{fig:OmegaX x/d=8.75}. 

\begin{figure}[b]
	\centering\includegraphics[width=0.6\linewidth]{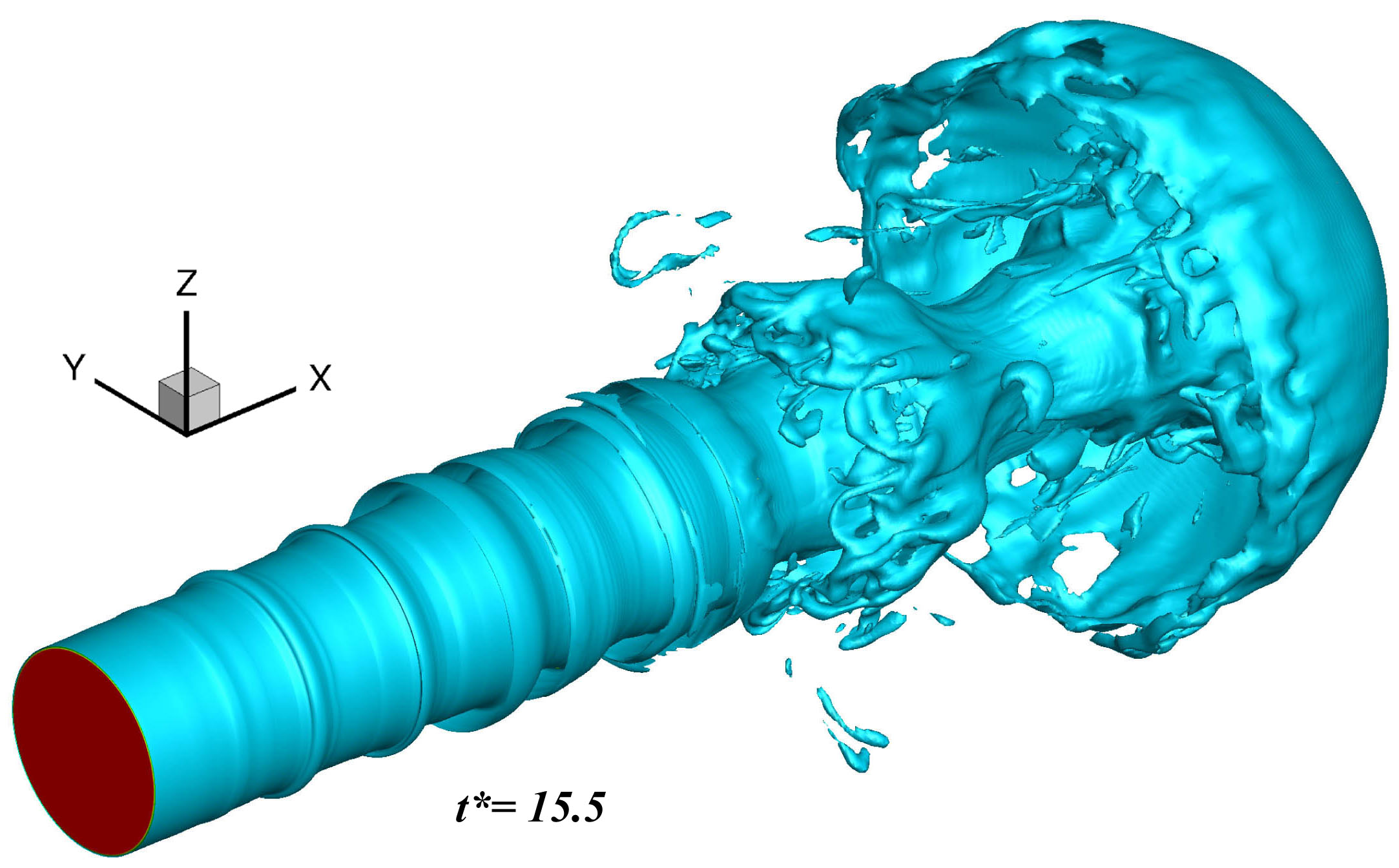}
	\caption{Liquid-jet surface at $t^*=15.5$; $\hat{U}=0.1$.}
	\label{fig:jet t15.5}
\end{figure}

Following the same wave at a later time ($t^*=12.5$), shown in Fig.~\ref{fig:jet t12.5}, it is seen that holes form on the lobes and they expand as the lobes get stretched in the axial direction. The liquid bridges that are formed around the lobe rim finally break and create the first ligaments, which then break into droplets or detach from the jet core. Figure \ref{fig:jet t12.5} also shows that there are in fact seven liquid lobes on each KH wave, as was inferred earlier by the number of counter-rotating axial vorticity pairs. Four of these lobes are already seen in this figure, and the other three are on the hidden side of the jet stem, which are blocked in this view. As we follow the jet structure at much later time ($t^*=15.5$ shown in Fig.~\ref{fig:jet t15.5}), we see that the same hole formation and breakup mechanism repeats for other waves as well. This confirms that the breakup mechanism on the jet stem in UR is periodic and occurs for all the waves formed in that range until they reach the BCR region and break into droplets and/or coalesce with the cap. This validates the temporal studies of Jarrahbashi et al.~\cite{ref:Dorrin1, ref:Dorrin2} and Zandian et al.~\cite{ref:Arash1, ref:Arash2}. The formation of holes on the rim of the jet cap (see Fig.~\ref{fig:jet t12.5}) is also conjectured to follow the same vortex overlapping mechanism, where the overlapping vortices in that case are the tip vortex and the downstream KH vortex (see Fig.~\ref{fig:vortex types}) as it runs along the inner side of the mushroom-shaped cap; however, these vortex structures are much harder to follow and are not analyzed in this study.

Based on $Re_l$ and $We_g$ values of the current study, this jet (without coaxial gas flow) should belong to Domain I as indicated by Zandian et al.~\cite{ref:Arash1}. In Domain I, lobes stretch directly into ligaments without formation of holes. The hole formation mechanism occurs in Domain II, at higher ranges of $Re_l$ and $We_g$. Thus, we can conclude from this study that addition of coaxial gas flow shifts the breakup mechanism, as this jet now belongs to Domain II, which is consistent with the use of relative velocity for defining $We_g$ and $Re_l$.

\begin{figure}[t]
	\centering\includegraphics[width=1.0\linewidth]{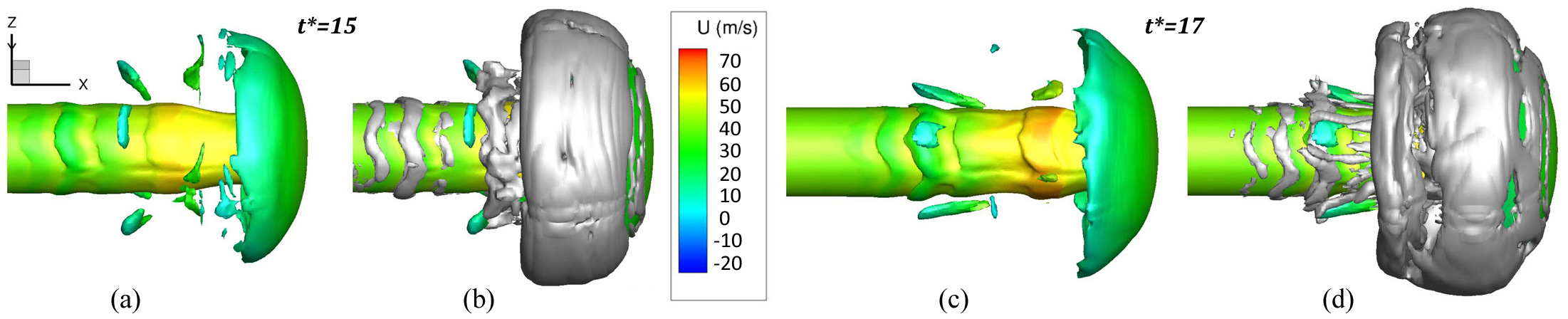}
	\caption{Liquid jet surface (a,c) and vortex structures (b,d) at $t^*=15$ (a,b) and $t^*=17$ (c,d); $\hat{U}=0.5$. The surface of liquid jet is colored by the axial velocity contours.}
	\label{fig:jet Uratio05}
\end{figure}

The same jet with a higher velocity ratio is simulated and the results are depicted in Fig.~\ref{fig:jet Uratio05} for $t^*=15$ and $17$. A few main differences are observed at a first glance between these results and the lower $\hat{U}$ cases. First, the azimuthal mode number has significantly decreased from seven (for $\hat{U}=0.1$) to four. Four liquid lobes are seen in this figure -- one in the front view, one on top, one on bottom, and one on the hind view which cannot be seen here, following the four axial pairs of KH vortices shown in Fig.~\ref{fig:jet Uratio05}(b,d). The next main change is that the lobes, ligaments and cap rim seem much thicker compared to lower $\hat{U}$. Because of this thickening in the lobe structures, the lobes do not thin easily and they are  stretched directly into thick ligaments. This means that the breakup mechanism has moved from Domain II towards Domain I by increasing $\hat{U}$. This clearly shows that the Reynolds and Weber numbers in such coaxial flow should be based on the relative gas-liquid velocity rather than just liquid jet velocity. By increasing $\hat{U}$, the relative velocity $U_r$ decreases, and thus, $Re_{l,r}$ and $We_{g,r}$ decrease too. This decrease in the Reynolds and Weber numbers is consistent with shifting from the hole formation breakup (Domain II) to lobe stretching mechanism (Domain I), as predicted by Zandian et al.~\cite{ref:Arash1}. Therefore, a more thorough analysis of the effects of velocity ratio is required to generalize the breakup mechanisms formerly developed for non-coaxial jet flows. This is left for a later study.

\section{Summary and Conclusions}

A three-dimensional round liquid jet with coaxial, outer gas flow is numerically analyzed. The evolution of instabilities on the liquid-gas interface were observed to be correlated with the vortex interactions around the liquid-gas interface using a $\lambda_2$ analysis. Two main regions were defined on the liquid jet separated by a large indentation on the jet stem with distinguished surface deformations. The Behind the Cap Region (BCR) is encapsulated inside the recirculation zone behind the mushroom-shaped cap. The KH waves formed on the jet core roll downstream in BCR and flow downstream until they coalesce with the cap. The vortices and surface deformations in BCR are not periodic and are controlled by the dynamics of vortices in the recirculation zone. The second region (Upstream Region, UR) is farther upstream of the cap. The gas speed is lower than the liquid jet in UR, and the shear caused by this upstream flowing gas stream relative to the liquid triggers a KH instability. The 3D deformation of the initially axisymmetric KH vortices leads to several liquid lobes. The lobes either thin and form holes at lower velocity ratios or stretch directly into elongated ligaments at higher velocity ratios, which could be explained by the vortex interactions in the UR region. The deformations developed in UR can be portrayed better in a frame moving with the convective velocity of the liquid jet with periodic conditions. The azimuthal and axial wavelengths of the instabilities and the breakup mechanism in UR can be well defined using a Reynolds and Weber number based on the relative gas-liquid velocity.%The gas-to-liquid velocity ratio is proven to be very important in defining the breakup mechanism as well as the azimuthal and axial wavelengths of the instabilities.

\section{Acknowledgements}
Access to the XSEDE supercomputer under Allocation CTS170036 and to the UCI HPC cluster were very valuable in performing our high resolution computations.

\end{document}